# Low-phase spectral reflectance and equivalent "geometric albedo" of meteorites powders

P. Beck[1,2], B. Schmitt[1], S. Potin[1], A. Pommerol[3], O. Brissaud[1]

[1]Institut de Planetologie et d'Astrophysique de Grenoble, UGA-CNRS
[2]Institut Universitaire de France, Paris, France
[3]Physikalisches Institute, Universität Bern, Sidlerstrasse 5, CH-3012 Bern, Switzerland

**Abstract:** Generally, the reflectance of a particulate surface depends on the phase angle at which it is observed. This is true for laboratory measurements on powders of natural materials as well as remote observations of Solar System surfaces. Here, we measured the dependences of reflectance spectra with phase angles, of a suite of 72 meteorites in the 400-2600 nm range. The 10-30° phase angle range is investigated in order to study the contribution of Shadow Hiding Opposition Effect (SHOE) to the phase behavior. The behavior is then extrapolated to phase angle of 0° using a polynomial fit, in order to provide grounds for comparison across meteorite groups (enabling to remove the contribution of shadows to reflectance) as well as to provide "equivalent albedo" values that should be comparable to geometric albedo values derived for small bodies. We find a general behavior of increasing strength of the SHOE with lower reflectance values (whether between samples or for a given samples with absorption features). This trend provides a first order way to correct any reflectance spectra of meteorite powders measured under standard conditions (g=30°) from the contribution of shadows. The g=0° calculated reflectance and equivalent albedos are then compared to typical values of albedos for main-belt asteroids. This reveals that among carbonaceous chondrites only Tagish Lake group, CI, and CM chondrites have equivalent albedo compatible with C- and D-type asteroids. On the other hand equivalent albedo derived with CO, CR and CK chondrites are compatible with L- and K-type asteroids. The equivalent albedo derived for ordinary chondrites is related to petrographic types, with low-grade petrographic type (type 3.6 and less) being generally darker that higher petrographic types. This works provides a framework for further understanding of the asteroids meteorite linkage in particular when combining with colors and spectroscopy.

1. Introduction

Solar System small bodies are a make-up of heterogeneous objects in term of size, orbit and colors. Observations available are much more numerous and detailed for the largest members, while for the smallest - and faintest - objects, orbit is generally the only available information. Still, there are 10 000s of objects for which, size, colors and albedo have been determined in addition to orbit (Izevic et al., 2001, Mainzer et al., 2011, Usui et al., 2019). Albedo is then one of the properties that is known for the largest number of small bodies and, combined with colors, could in principle be used to help mapping the composition distribution of small bodies across the solar system. However, the interpretation of albedo in term of composition is degenerate; the brightness of a material can be related to chemistry (the presence of Fe-related absorption features, the presence of opaque phases, metals, oxides, organics,...) as well as to physical properties such as grain size. Still, albedo has been used as a zero order connection between asteroid classes to meteorites groups by qualitatively comparing albedos to reflectance levels measured on meteorites: the dark asteroid families are often named C-complex, where C refers to "carbonaceous", by analogy to the dark carbonaceous chondrites. This has so far remained a qualitative comparison since reflectance levels cannot be compared to albedos: they are not the same physical quantity by definition, which forfeits a direct comparison.

There are multiple types and definitions of albedo. In the case of large asteroid surveys, the geometric albedo in the visible, $p_v$, is determined on the basis of mid-infrared observations coupled with a thermal model and visible observations (Lebofsky et al., 1986). The geometric albedo is defined as the brightness ratio between the reflected sunlight (averaged spectrally over the visible range) to that of a perfectly lambertian disk with the same cross-section, everything being observed at a phase angle of g=0°. This quantity is therefore different from laboratory measurements obtained on meteorite samples, which typically provide biconical or bidirectional reflectance spectra in a phase angle range around 30°. While the definition of $p_v$ is at zero phase angle, note that asteroid observations are essentially conducted at low phase angles but only rarely obtained at exactly g=0°. They are extrapolated to g=0° using measurements (or guesses) of the phase coefficient (the dependence of brightness on phase angle). This approach is complicated by the fact that the opposition effect, the strong increase of reflectance towards 0° phase, shows variability in both its magnitude and shape depending on the composition and physical properties of the surface material.

Here, we present an attempt to provide laboratory measurements that can in principle be used to compare with the geometric albedo of a smooth-surfaced asteroid, covered by particulate meteorite-like material. Previous investigations have looked at some aspects of this topic on extra-terrestrial samples (Gradie et al., 1980; Gradie and Veverka, 1986; Capaccioni et al., 1990), but the originality of our approach is to focus on the low phase angle range and to sample a large diversity of samples. For that we present multispectral measurements on a suite of 72 meteorite samples belonging to 10 different classes, with a special emphasis on carbonaceous chondrites. For each sample, multispectral measurements were obtained in the 400-2500 nm spectral range, at phase angles of 10, 20 and 30°. This enables to extrapolate the reflectance spectra at 0° phase angle for each sample, and to calculate the reflectance at 0° phase angle averaged over the Solar spectrum, two quantities that can be seen as equivalent to $p_v$ and p. We then compare these values to albedo values derived for various asteroids classes, and discuss the connection between meteorite classes and asteroid families.

## 2. Samples & Method

2.1 Samples

A suite of 72 meteorite samples was analyzed for this study. We particularly focused on carbonaceous chondrites, which have been related in the past to C-type asteroids. Samples from the CI, CM, CR, CO and CV groups were analyzed together with ungrouped C2 chondrites. Many of these samples were provided through the Antarctic meteorite research program. Meteorites were grinded in an agate mortar, in order to obtain a particulate sample. Previous samples prepared with the same protocol resulted in typical grain size in the range 30-200 µm (Beck et al., 2012). A list of samples used in this study and their reflectance values is provided in supplementary materials.

2.2 Bidirectional reflectance measurements

The reflectance measurements were obtained with the SHADOWS spectro-gonio radiometer (Potin et al., 2018). For each sample, we measured the reflectance between 400 nm and 2500 nm, every 100 nm at a spectral resolution of around 4 nm in the 360-670 nm range, 8 nm in the 680-1420 nm range and 16 nm in the 1430-2640 nm range (Figure 1). This wavelength range was measured in order to cover a significant fraction (88%) of the solar energy spectral distribution.

The SHADOWS instrument enables to measure bi-directional reflectance over a range of incidence and emergence angles, down to phase angles of 5°. For this work spectra of each sample were measured at 10, 20 and 30° phase angle for nadir incidence (0°). The instrument was used in standard mode with a spot size of 5.2 mm (much larger than the grain size) and the typical sample mass used was around 50 mg. Spectra were obtained under ambient conditions and calibrated using Spectralon™ and infragold™. The system used a 20mm diameter diaphragm placed in front of the optics of both detectors to increase the bidirectionality (i.e. the angular resolution) of the observation to ±0.8° (therefore the angular resolution of the system is of 3°, constrained by incidence).

For each sample, the reflectance values obtained are extrapolated to zero phase angle using a second order polynomial fit of the measurements at 10, 20 and 30°. This is possible in the 400-1000 nm range measured by a silicon detector with a high signal to noise ratio (SNR). This is not possible in the near-IR range where the SNR of the InSb detector was not sufficient to enable a reliable extrapolation to zero-phase. The significance of the 3-points polynomial approach is investigated in the case of four samples (1 CO, 1 CV, 1 CR, and one ordinary chondrite) by measuring the phase-curve every two degrees from g=8 to g=40° (Fig. 2) and comparing the interpolated values to g=0° using either the 3-point approach or the full dataset (second-order polynomial fit of the 17 data points). This test reveals that using 3 points or 17 points does not change the zero-phase interpolated value by more than a few percent (<0.005 in absolute reflectance) (Fig. 2).

Astronomical observations of small bodies, and derivation of zero-phase magnitude from phase curves are generally performed by using an exponential fit to the magnitude vs. phase angle relation. We chose not to use the same analytical formalism because these observations are not directly analogous to our laboratory measurements. Indeed the phase curves of small bodies are a combination of photometric behavior of the surface, and the fraction of the objet that is illuminated when seen from the observer. As a consequence, the polynomial approach was used since it provided a simple and satisfactory fits to the 17 points curves.

## 3. Calculation of « albedo » and zero phase reflectance.

3.1 Method

At present, the SHADOWS instrument does not allow to perform measurement at zero phase angle. Therefore, for each wavelength the reflectance at zero phase angle is obtained by extrapolating the reflectance at 10°, 20°, 30° to a phase of 0° using a second order polynomial. At low phase angle (g<30°), two contributions result in an increase of reflectance with decreasing phase angle (Hapke, 2012). First, the shadow-hiding opposition effect (SHOE) that typically occurs below 30° is related to the fact that at lower phase, there are less shadows at grain scale, and then the surface is brighter. It is a geometrical effect although its magnitude may depend on the scattering properties of the grains. A second contribution is the coherent back-scattering opposition effect (CBOE), which is related to additive or destructive interactions of photons and occurs for phase angles typically below 2° for many particulate samples and for the lunar surface (Hapke, 2012).

In the case of small bodies observations, the geometric albedo that was computed is based on thermal modeling of mid-IR thermal emission observations as well as on the values of the visible absolute magnitude (H) and phase slope parameter (G). The latter is measured from telescopic observations that are only rarely obtained at very low phase angle (Harris, 1989) and therefore may not take into account the CBOE. This is why we prefer in this work not to use zero-phase angle measurements directly but rather to use an interpolation approach.

Two quantities are calculated that could in principle be used for comparison with small bodies observations. The first one is the 550 nm reflectance at zero-phase angle, which is calculated by averaging the 500 and 600 nm zero-phase reflectances, that are estimated using the polynomial interpolation. This quantity can in principle be compared to the visible geometric albedo $p_v$ and may be called *calculated zero-phase reflectance* or *equivalent visible geometric albedo*.

A second quantity that is calculated is the solar-spectrum averaged zero-phase reflectance. For each sample and each phase angle, the spectral integral of a black body spectrum at 5500°C multiplied by the reflectance spectrum of the sample is divided by the spectral integral of the same blackbody multiplied by the spectrum of a lambertian surface (reflectance of 1).

$$A_{Lab}^g = \frac{\int_{\lambda=400}^{2500} R_\lambda \times BB_\lambda d\lambda}{\int_{\lambda=400}^{2500} BB_\lambda d\lambda} \qquad (eq.\ 1)$$

This calculated quantity represents the ratio between the solar radiation reflected by the sample surface at a given phase angle to that reflected by a lambertian surface. This quantity was calculated for g=10, 20 and 30° and then interpolated to g=0° using a second order polynomial fit, in order to estimate the fraction of solar radiation reflected by the surface at zero-phase angle. This quantity is somehow analogous to the geometric albedo $p$ and may be called *equivalent geometric albedo*.

3.2 Cautions and underlying hypothesis

When comparing calculated zero-phase reflectance and equivalent geometric albedo from our work to asteroid observations, there are some strong underlying hypotheses that the reader needs to keep in mind:

- Photometric properties are hypothesized to be dependent on phase angle only, while in reality incidence and emergence angles also have an independent role (see for instance Beck et al., 2012 or Potin et al., 2019), an observation also noted at larger phase angles and having led to the weighting of measurements made at high incidence angles in the photometric inversion procedures (e.g., Souchon et al., 2011).

- The contribution from the CBOE is not taken into account. The zero-phase reflectance derived here cannot be used to compare with observations obtained at very-low phase angles (<2°). For low-albedo samples (i.e. carbonaceous chondrites) the CBOE might be moderate (Skhuratov and Helfenstein, 2001) but detailed investigations of CBOE on carbonaceous chondrites are not available at present.

- The impact of large-scale shadows is not taken into account when comparing disk-integrated observation to laboratory measurements. The effects of large-scale shadows are expected to be modest at low phase making this hypothesis reasonable (Hapke et al., 2012).

- A particulate material with a wide range of particle sizes is covering asteroids surfaces. The measurements in this work were obtained on fine powders and the behavior of a surface covered by large particles (>1 cm) could differ from that of a surface covered by smaller ones (especially for weakly absorbing materials). Grain shape as well as the style of topography at all scales may also play a role (e.g., Shepard and Campbell, 1998; Helfenstein and Shepard, 1999; Cord et al., 2003; Skhuratov et al., 2005; Souchon et al., 2011).

- We did not carry out any measurement on pure Fe-metal, so this work does not apply to metal dominated surfaces.

## 4. The scattering behavior of meteorites at low phase angle

4.1 A back-scattering behavior at low phase angle

All 72 meteorite samples analyzed in this study reveal a similar behavior of increasing reflectance at low phase angle. We quantify the relative magnitude of the opposition effect as the ratio between the extrapolated reflectance at zero phase angle and the reflectance in standard geometry (i=0°, e=30°). Note that this definition is close but slightly different from that in Beck et al. (2012) (i=3°, e=30°). In figure 3, the relative intensity of the opposition effect is shown as a function of the reflectance at standard geometry for the six wavelengths measured below 1 μm (zero-phase angle reflectance could not be extrapolated for the near-IR range due to the lower SNR). This graph reveals a general behavior among all samples and wavelengths, namely that the relative magnitude of the opposition effect is greater for darker surfaces. The values found in this work are in fair agreement with previous results we obtained on 6 samples (Beck et al., 2012, Fig. 3,5).

For asteroid observations this translates into the fact that darker asteroids should present higher phase coefficient values, which is indeed the case (Belskaya and Shevchenko, 2000; Longobardo et al., 2014, 2016). This increased back-scattering behavior for low-reflectance surfaces can be explained by the shadow-hiding opposition effect (SHOE). Indeed darker materials have a steeper phase slope due to a combination of surface texture and radiative transfer effects. When grain sizes are of the order of the surface roughness length scale, multiple scattering dilutes the effect of shadowing for transparent materials (i.e. bright surfaces). In the case of more opaque materials, the effect of shadowing is more pronounced compared to the reduced multiple scattering contribution (Skhuratov et al., 2005).

Several analytical functions were tested in order to model the distribution observed in figure 3. Among those investigated, the function that resulted in the best quality of fit is a log-normal distribution, that is expressed as:

$$\frac{R_{\lambda,g=0°}}{R_{\lambda,g=30°}} = C_1 + C_2 \exp\left(-\left[\frac{\ln\left(\frac{R_{\lambda,g=30°}}{C_3}\right)}{C_4}\right]^2\right) \quad \text{(eq. 2)}$$

with $C_1$=5.48 ± 1.14, $C_2$=-4.40 ± 1.14, $C_3$=0.388 ± 0.019, $C_4$=6.08 ± 0.95

This law should in principle enable to calculate a low phase reflectance from a reflectance value measured under standard conditions. Note that this law does not take into account the CBOE, and should not be used to compare laboratory data to observations at very low-phase angle (g<2°) where the CBOE starts to play a significant role. Note also that this law was obtained for powders and should not be applied to bulk samples (i.e. not powdered) until it is proven valid for such samples as well (which is most likely not the case). law also only applies to reflectance lower than 0.5 which is the range investigated here (this is fortuitous, only related to the arbitrary choice of a mathematical description of this dataset, which is not a bijection in the range of value investigated here). Conversely, this law may enable to convert asteroid albedo values into reflectance under standard laboratory conditions, for comparison with laboratory data measured under "standard" conditions.

4.2 The low-phase relative blueing

The behavior observed in figure 3 could also be modeled by a law f(x)=x/(a+x), but with a slightly lower quality of the fit. This means somehow that the shadow-hiding opposition effect studied here is close to being an additive contribution:

$$R_{\lambda, g=0°} \approx C + R_{\lambda, g=30°} \text{ with C a constant value}$$

This means that in relative intensity the effect will appear stronger for darker samples. Another consequence is that when working in relative reflectance, which is a common way to analyze ground-based observations, phase angle can have an effect on the calculated spectral slope. To illustrate this point, the spectra obtained for 4 meteorite samples at different phase angles are normalized at 500 nm and displayed in figure 1. These graphs reveal that a red sample will appear bluer at low phase angle if the slope is calculated after spectral normalization, and particularly in the case of dark meteorites (see the CV chondrite LAP 02206 in figure 1). Such a behavior is also revealed when the visible slope is calculated for each meteorite in absolute and relative reflectance, at g=0 and g=30° (Figure 4). This is due to the fact that the SHOE behaves almost like an additive contribution, because it is only due to external reflections on the grains, therefore without absorption. This also means that at low-phase a spectrally flat sample will not become blue, but that a sample with a spectral slope will have a decreasing spectral slope at decreasing phase angle, if the slope is calculated in normalized reflectance. The redder the spectra, the more pronounced the low-phase blueing (Figure 4). Note that, at least in the case of S-type asteroids, these effects appear minor when compared to the magnitude of space weathering effects (Vernazza et al., 2009).

4.3 The opposition effect as a function of meteorites groups in carbonaceous chondrites

In figure 5, the same data as figure 3 are shown but each meteorite group is color coded, in order to investigate differences in magnitudes of SHOE among the various groups investigated. Within that plot, meteorites groups with particular high opposition effect will appear above the average trend (the global fit presented in Figure 3) while meteorites groups with relatively low opposition effect will appear below. This graph reveals that among carbonaceous chondrites, some CV chondrites appear to have an unusually pronounced SHOE. On the other hand heavily aqueously altered meteorites (C2, CI and CM) appear to have a relatively less pronounced SHOE, and meteorites from the CO and CR groups appear to be in between.

The existence of these group behaviors may be related to different mineralogy (optical properties of grains), grain sizes and grain shapes. It is certainly difficult to disentangle which property is responsible for these contrasted behaviors. The first observation that should be made is that meteorites groups that experienced the strongest aqueous alteration tend to lie below the global fit. Aqueous alteration has mineralogical consequences, since it induces the hydrolysis of primary phases and the production of phyllosilicates and secondary opaques. Aqueous alteration has also physical consequences, for instance to increase porosity (Britt and Consolmagno, 2000) or decrease strength of the samples (personal experience of the authors with grinding such samples).

The presence of a significant amount of metal may produce a peculiar behavior, but metal abundance varies in the order CM-C2-CI < CV < CO < CR (Krot et al., 2006). Also note that the only metal rich meteorites studied here (mesosiderite) tend to show a rather low SHOE. The presence of an elevated amount of CAI in CV could provide a lead to explain the unusually strong opposition effect observed, but CAI are abundant in CO as well.

## 5. Low-phase reflectance and equivalent geometric albedos among meteorite families

5.1 Comparison to RELAB data

Several parameters may in principle impact the reflectance value measured for a given sample. These include grain size, sample heterogeneity (meteorites are often breccia), as well as measurement uncertainty. Also, because the final goal of this work is to provide a way to compare laboratory data to observations, it is important to assess how laboratory measurements may differ between different laboratories. In order to investigate interlaboratory variability, we chose to compare to data from the RELAB facility since it provides at the moment the most extensive suite of measurements on extra-terrestrial materials. In figure 6 we present the range of reflectance values at 550 nm for different classes of meteorites using the IPAG setup (SHADOWS, this study and Eschrig et al., submitted) and data obtained on meteorite powders with the RELAB setup (Brown University). This graph shows a good agreement for all classes of meteorite studied here. The only difference found is for type 3 ordinary chondrites (OC), which appear to show a globally lower reflectance in the dataset measured with SHADOWS, but this difference may be related to the different definition of the petrographic types used at IPAG (based on Raman studies, Bonal et al. 2018).

5.2 Equivalent albedo values

5.2.1 Carbonaceous chondrites

The equivalent albedo values derived for our suite of meteorites vary significantly from family to family (Fig. 7). In the case of carbonaceous chondrites, the average equivalent albedo decreases in the order CK > CR > CV > CO > CM > CI > TL. A first explanation for this variation could be the different amount of carbonaceous compounds in those meteorite samples. In figure 7, the average equivalent albedo per meteorite class is plotted against the average carbon content. The carbon content was calculated using data in Alexander et al. (2012, 2013) as well as Jarosewich (1990) and Pearson et al. (2006). As can be seen in Fig. 7 a rough correlation exists between carbon content and equivalent albedo but correlation does not imply causality. Indeed, the carbon content in carbonaceous chondrites is a proxy for the amount of matrix and this correlation might also be explained by the fact that meteorites rich in carbon also tend to have an elevated amount of dark matrix, which is enriched in iron-rich opaque minerals.

### 5.2.2 The impact of thermal processing on carbonaceous chondrites

A significant fraction of CM chondrites has experienced a post-aqueous alteration heating process (Nakamura et al., 2005). The measurements obtained here on heated CM chondrites enable to assess the impact of thermal processing on equivalent albedo of primitive carbonaceous chondrites. Results show that on average heated CM chondrites are not darker than more primitive CM chondrites, but they show more variability in equivalent albedo values (see supplementary table). This variability was explained in Beck et al. (2018) by the fact that upon heating CM chondrites first become darker due to modifications of organic compounds, and then become brighter upon more intense thermal processing, as a consequence of the recrystallization of the matrix.

### 5.2.3 Ordinary chondrites

Our dataset contains mostly ordinary chondrites of low petrographic types since we focus here on "primitive meteorite samples". It is therefore important to remark that this dataset is not representative of the average of the whole range of metamorphic grades of ordinary chondrites.

We find that primitive ordinary chondrites of petrologic type 3, with an average $p$ of 0.154 +/-0.024 (1 sigma, n=6), appear darker than petrologic types > 4. This difference can be explained by the fact that small opaque phases that are present in low petrographic grade meteorites tend to coalesce during the metamorphic process, and becomes less efficient in darkening the sample. Low-metamorphic grade ordinary chondrites have a higher equivalent albedo than CI and CM chondrites, but their equivalent albedo values are in the range of CO, CV and CR chondrites.

### 5.2.4 HED meteorites

Values derived for HED meteorites vary typically in the 0.2-0.5 range (see supplementary table). Within this suite the howardite sample has the lowest albedo when compared to eucrite and diogenite but the dataset is too small to address whether this is a general behavior.

5.3 p vs pv

When asteroid albedo values are determined, an assumption often made is that p/pv=1 (Lebofsky et al., 1986). In Figure 8, we present the values of the ratio of equivalent albedos to equivalent visible albedos, which can be considered as analogous to p/pv. The values found range from 0.85 to 1.25 with an average of 1.04. The values are mostly above 1 except for HED meteorites and type 4-5 ordinary chondrite. The specificity of these meteorites is the presence of a strong absorption band around 850 nm that absorbs a significant fraction of the solar energy. On the other hand, most samples studied have a red slope, which may explain the overall values above 1.

## 6. Comparison to small bodies

In figure 9 we present a comparison of equivalent albedo values derived for our series of meteorites, to the asteroid family modal distribution albedos as obtained by DeMeo and Carry (2013). The DeMeo and Carry (2013) data were obtained by mapping DeMeo et al. (2009) spectral endmembers on multi-filter sky surveys, enabling to study objects down to diameters as low as 5 km and providing the most detailed taxonomy of main-belt objects. The values used for comparison are the modes of the albedo value distribution for each taxonomic group (Table 4 in DeMeo and Carry, 2013,).

6.1 "Carbonaceous" asteroids vs carbonaceous chondrites

While the denomination "C"-type asteroid initially stands for carbonaceous, a first important observation is that only a fraction of carbonaceous chondrites appears compatible with C-type asteroids. The CO, CV, CR and CK chondrites have albedos significantly higher than CM chondrites and are too bright to be related to C-type. These meteorites classes represent about 50% of the carbonaceous chondrite falls and their possible parent bodies will be discussed in sections 6.3-6.4 and 6.5.

Among C-type asteroids, a significant fraction (⅓ to ½) is hydrated and shares very strong spectral similarities with CM chondrites (Vernazza et al., 2015, 2016; Vernazza and Beck, 2017). The average albedo value derived for CM chondrite (including heated samples) is 0.066 ± 0.019 which is in good agreement with C-complex asteroid (0.054 ± 0.023, DeMeo and Carry, 2013; Fig. 9). This good agreement reinforces the connection between hydrated C-complex asteroids (Ch, Cgh) and CM chondrites.

The rest of the C-complex that do not show evidences of hydration remains puzzling and there are at present two competing hypothesis. The first one is that they represent thermally metamorphosed CM chondrites (Hiroi et al., 1993). From an albedo perspective this hypothesis remains valid since heated CM chondrites have albedo values consistent with C-type asteroids (Table 1) except for those that have been the most extensively thermally processed (Beck et al., 2018). However, this hypothesis is challenged by the fact that heated CM chondrites do show evidence of hydration at 3-µm (Garenne et al., 2016; Beck et al., 2018) and they often show olivine feature in their mid-infrared spectra (Beck et al., 2014) unlike "non-hydrated" C-type (Vernazza et al., 2015). An alternative interpretation is that these objects may represent loosely consolidated material related to Interplanetary Dust Particles (Vernazza et al., 2015).

6.2 S-type: the scarcity of low-metamorphic grade objects

The albedo values calculated for type 3 ordinary chondrites are much lower than those obtained for the LL4-5 chondrites, as well as values found for type 4 and above ordinary chondrites measured under low phase angles (Capaccioni et al., 1990). Type 3 ordinary chondrites escaped the thermal metamorphism event experienced by most ordinary chondrites (Bonal et al., 2016). The thermal metamorphism process is explained by the build-up of heat within the parent body due to decay of short and long-lived radionuclides, and type 3 ordinary chondrites are thus expected to have been at some point the surficial lithology of their parent bodies.

The albedo values of S-type (0.247 ± 0.084, DeMeo and Carry, 2013) is significantly higher than values obtained for type 3 ordinary chondrites (Fig. 9). This difference cannot be attributed to space weathering that tends to darken asteroid surfaces with time

(Pieters and Noble., 2016). This observation suggests that the surface of S-type asteroids is not covered by material analogue to type 3 chondrites, but rather by more thermally processed material analogous to type 4 or above. Such an observations is in agreement with a fast accretion of S-type parent bodies, followed by fragmentation and surface brecciation (Vernazza et al., 2014).

6.3 K -and L-types

L-type asteroids are a peculiar class of objects with an absorption band above 2-µm, which has been explained by the presence of FeO bearing spinel (Burbine et al., 1991) in excess of what is observed in carbonaceous chondrites (Sunshine et al., 2008). This led to the idea that they could represent a reservoir of early solar system material not sampled by meteorites, enriched in refractory inclusions (Sunshine et al., 2008). Another peculiarity of L-type asteroids is the presence of unusual polarimetric curves (Devogèle et al., 2018; Cellino et al., 2005) with an inversion angle much larger than other asteroid taxonomic types (i.e. > 25°).

K-type asteroids are characterized by the presence of a red-slope and a moderate 1-µm silicate band (olivine). K-type asteroids are distributed among the entire main-belt according to multi-color surveys but a large number of K-type objects are found among the EOS family (Mothes-Diniz et al., 2008; DeMeo and Carry, 2013).

K-type objects can be distinguished from the L-type by the presence of a weaker 2 µm band and a stronger 1-µm band. The average albedos derived for K- and L-type by DeMeo and Carry (2013) are similar (so it is difficult to distinguish between the two types in multicolor surveys). Among the meteorites studied, only CO, CK and CV groups appear consistent with a connection to L and K-types based on their low-phase reflectance (Fig. 9); L- and K-type are thus very likely related to carbonaceous chondrites but not all types of carbonaceous chondrites.

Since albedo does not discriminate between CO, CK and CV, to further investigate the nature of L- and K-types, it is necessary to turn to spectroscopy. Spectroscopic surveys of the EOS family have revealed the presence of a strong 1-µm band but only weak 2-µm features (Mothes-Diniz et al., 2008). These spectral characteristics are typically found in laboratory spectra of CK meteorites, while spectra obtained for CO and CV tend to show a lot a variability from featureless spectra, to spectra containing both a 1- and a 2-µm band.

Therefore, some CO and CV chondrites have affinities with L-type asteroids but not all meteorites of these groups.

6.4 D-type vs Tagish Lake group

The Tagish Lake meteorite was the first identified as having a reflectance spectrum similar to D-type asteroids (Hiroi et al., 2011). Since then, a few other meteorites have been recognized as having spectra similar to Tagish Lake and D-types. From an albedo perspective, the connection between D-type and Tagish Lake appears reinforced (Fig. 9) given the similarities found in "albedo" values. The spectra similarity and the albedo similarity are therefore making a strong case for a meteorite-parent body relation, but there are some major difficulties in this connection. A first one is the difference in mid-IR emissivity spectra that was noticed by Vernazza et al. (2011), which suggests a very fine-grained surface (not lithified) similar to comets. The second one is the lack of a strong 3-µm band on the only D-type object visited by a spacecraft, comet 67P/CG (Cappaccioni et al., 2015; Quirico et al., 2016), while all Tagish Lake samples show the presence of abundant hydrated minerals, leading to a well-defined 3-µm band (Gilmour et al., 2019). Therefore Tagish Lake has strong affinities to D-type asteroids when looking at the VNIR, but its mineralogy appears to be different from D-type surface when looking at the 3- an 10-µm regions.

An explanation to reconcile these observations is that Tagish Lake arises from the interior of a D-type object where sufficient heat enabled to transform primary phases into phyllosilicates but without strongly altering its VNIR properties (and then its spectrum and low-phase reflectance). If the D-type population accreted late, $^{26}$Al decay was insufficient to induce a total differentiation of 100 km sized objects (Neveu and Vernazza, 2019), and while a muddy interior may have existed, their surface may have remained pristine. The remaining difficulty is to explain how aqueous alteration may have produced abundant phyllosciliates, without changing the VNIR spectra. One possibility is that the peculiar spectrum of Tagish Lake is related to the abundant macromolecular organics, (Herd et al., 2011) and that they were not modified during aqueous alteration. Experiments on soluble organics mixed with phyllosilicates under hydrothermal conditions reveal that they experience important transformations (Vinogradoff et al., 2018; Vinogradoff et al., 2020). In the case of more refractory organics (IOM, insoluble organic matter), the impact of hydrothermal alteration is less understood, in particular due to difficulties in generating a valuable laboratory analogue. An approach to investigate the impact of aqueous alteration on the D-type signature that can be pursued, is to look at the VNIR signature of different lithologies of Tagish Lake, that experienced different level of aqueous alteration. Such an approach was started in Gilmour et al. (2019), and showed that all lithologies investigated have a D-type signature, while they experienced different levels of aqueous alteration.

7. Summary and conclusions

The results and conclusions can be summarized as follows:

An approach is proposed using low-phase angle measurements of meteorites to provide observable quantities analogous to geometric albedo for meteorites powders. This is done under a number of hypotheses listed in 3.2.

There appears to be a general trend in the low-g behavior of meteorite powders. The increase of reflectance due to the shadow-hiding opposition effect (SHOE) is an almost additive contribution. A relation is derived to obtain the zero-phase angle reflectance from standard geometry observations (Fig. 3). Note that this relation (eq. 2) does not take into account the coherent back-scattering opposition effect.

Albedo values derived for Tagish-Lake lithologies and "Tagish-Lake group" meteorites are in agreement with D-type asteroids. We propose that they represent samples from the interior of a D-type or a piece of D-type asteroid that experienced heating and hydrothermal alteration.

Albedos values derived for CM chondrites are in good agreement with values found for C-type asteroids. Heated CM chondrites tend to show more variability in albedo with both brighter and darker samples when compared to "standard" CM chondrites. This reinforces the relation between CM chondrites and hydrated C-complex asteroids.

Carbonaceous chondrites from the CO, CV, CR and CK groups are brighter than those from the CI, CM and Tagish Lake groups. The derived albedo values are higher than for C-type asteroids, while in the range of K and L-type. This confirms that a significant fraction of carbonaceous chondrites does not originate from C-type. Combined with spectroscopic observations, this also reinforces the relation between K-type (mostly EOS family) and CK meteorites. This dataset further delves into the mystery of the parent bodies of CO-CV-CR meteorites for which there are no well-identified parent bodies that match the spectra and albedo of these meteorites.

## Acknowledgements

This work was funded the European Research Council under the H2020 framework program/ERC grant agreement no. 771691 (Solarys). Comments by Patrick Pinet greatly improved the manuscript. Additional support by the Progamme National de Planétologie and the Centre National d'Etude Spatiale is acknowledged.

***

Figure captions

Figure 1: Multispectral observations of powdered meteorites (LL4, CO, CR, CV) at phase angles ranging from 8 to 40° (incidence=0°). Spectra are shown in reflectance factor (left graphs) or in relative reflectance normalized at 500 nm (right graphs). The black line is the extrapolated reflectance at g=0° using a polynomial fit of the data (see Fig. 2).

Figure 2: Examples of fits of the phase curve for the extrapolation of the reflectance at g=0° using a 2$^{nd}$ order polynomial. The graphs show the reflectance at 500 nm as a function of phase angle for the same samples as in figure 1 and compare the extrapolated value at g=0° using a 3-point interpolation (all 72 samples from this study)

against a 17-points interpolation (tested only for these 4 samples). The difference between the two extrapolated values at g=0° is less than 0.005.

Figure 3: « Relative magnitude » of the opposition effect as a function of the reflectance at standard geometry. This magnitude is defined as the ratio between the extrapolated reflectance at g=0° and the reflectance in standard geometry.

Figure 4: The effect of phase angle (g=0 or g=30°) on the 400-900 nm visible spectral slope (in µm$^{-1}$)   whether calculated in reflectance or in relative reflectance. In the case of reflectance, a slight redening is present for low-phase angle. When slope are calculated for relative reflectance sectra, a significant blueing is observed at low phase angle for the spectra with initially red slopes.

Figure 5: « Relative magnitude » of the opposition effect as a function of the reflectance measured under standard geometry. This graph shows the same data as figure 3, but color- and symbol-coded according to meteorite groups. The data from Beck et al. (2012) is also shown (circles) with the same color code.

Figure 6: Comparison of the reflectance at 550 nm for meteorite powders of carbonaceous chondrites measured at IPAG and at RELAB for different meteorite groups. This graph shows that while some variability is present for each meteorite group, the range of value is in agreement for measurements obtained in two different laboratories.

Figure 7 : Relation between meteorite equivalent albedo and average carbon content for different meteorite groups. Carbon contents are from Alexander et al. (2012), Jarosewich (1990) and Pearson et al. (2006)

Figure 8: Ratio between equivalent geometric albedos and equivalent visible geometric albedos for different meteorites belonging to different classes.

Figure 9: Comparison of the reflectance at 550 nm from this work for different meteorites groups, the calculated equivalent geometric albedo, and geometric albedos of asteroids for different taxonomical families (DeMeo and Carry, 2013).  Top: linear scale. Bottom: logarithmic scale.

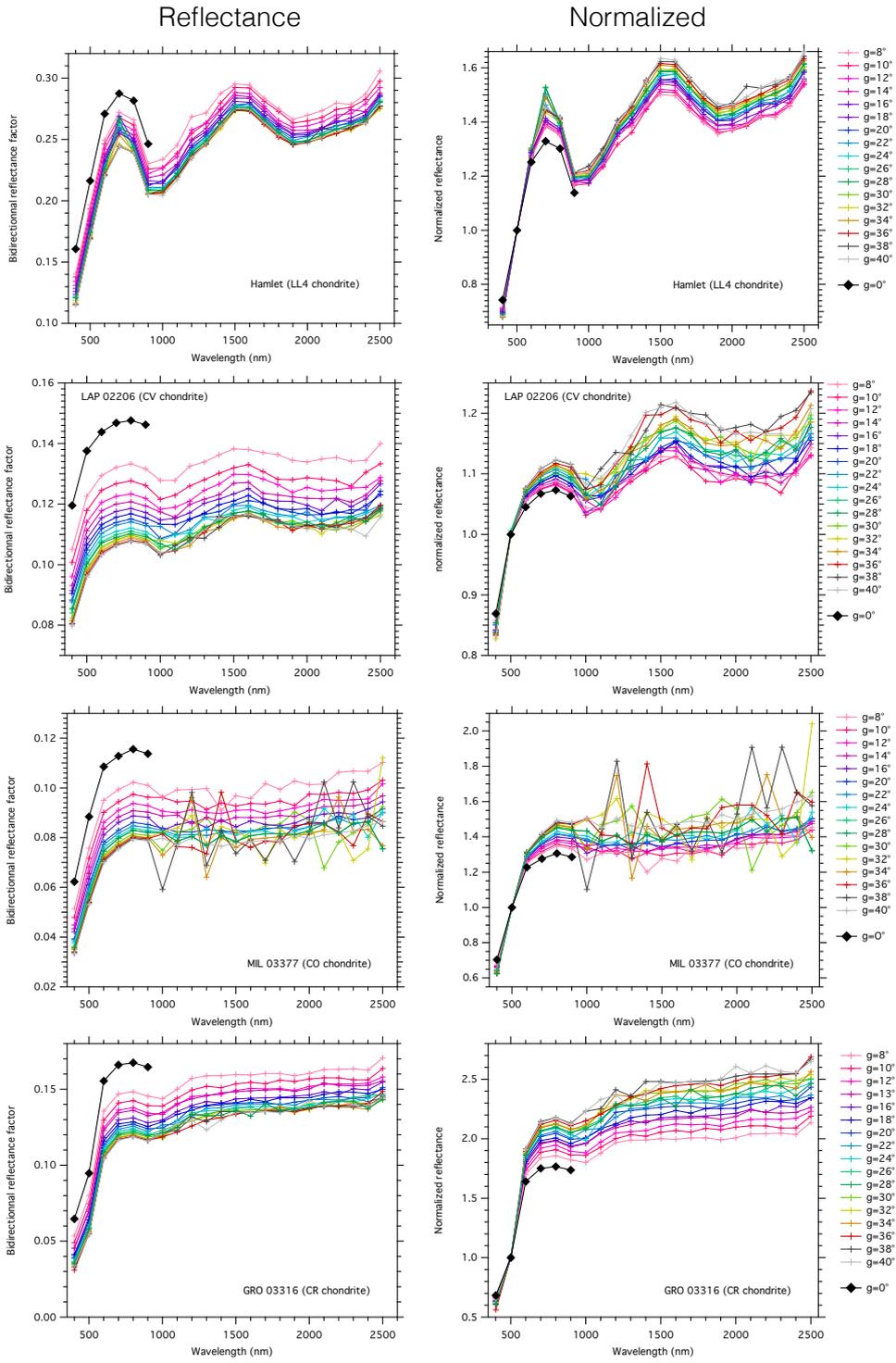

FIGURE 1

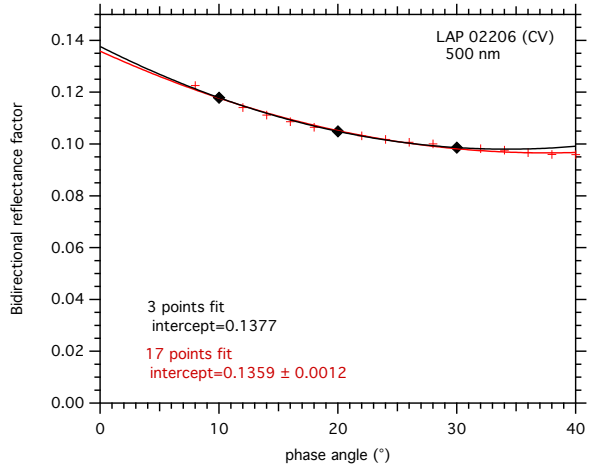
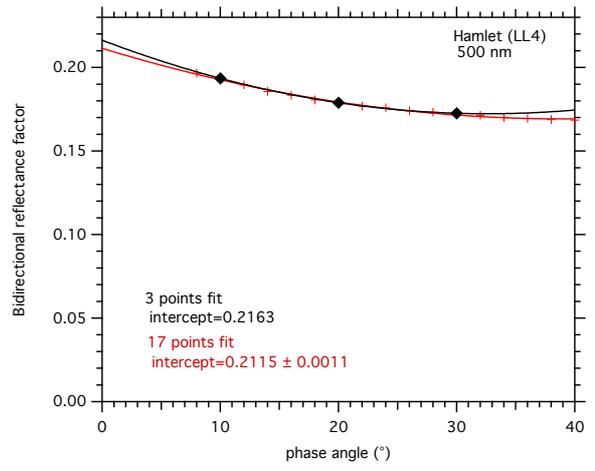
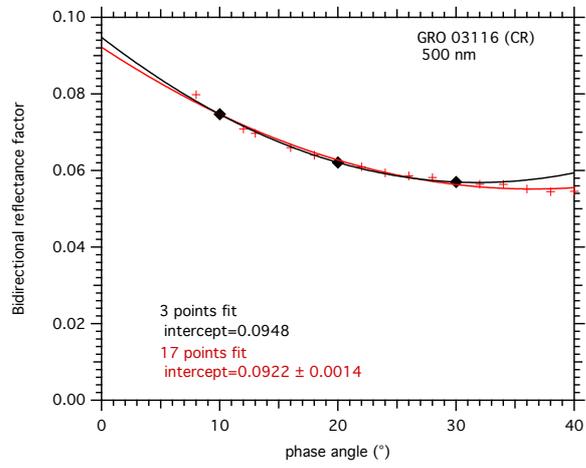
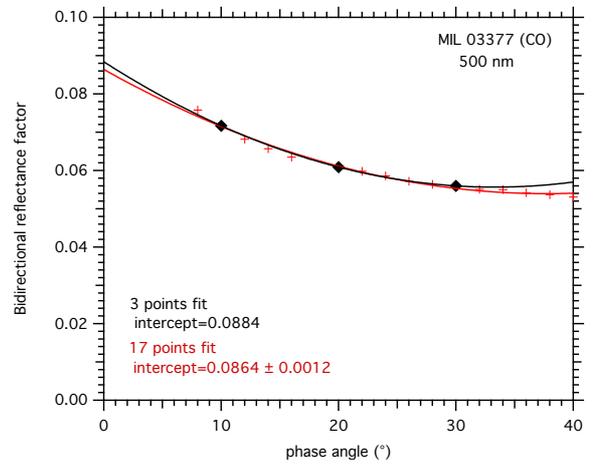

FIGURE 2

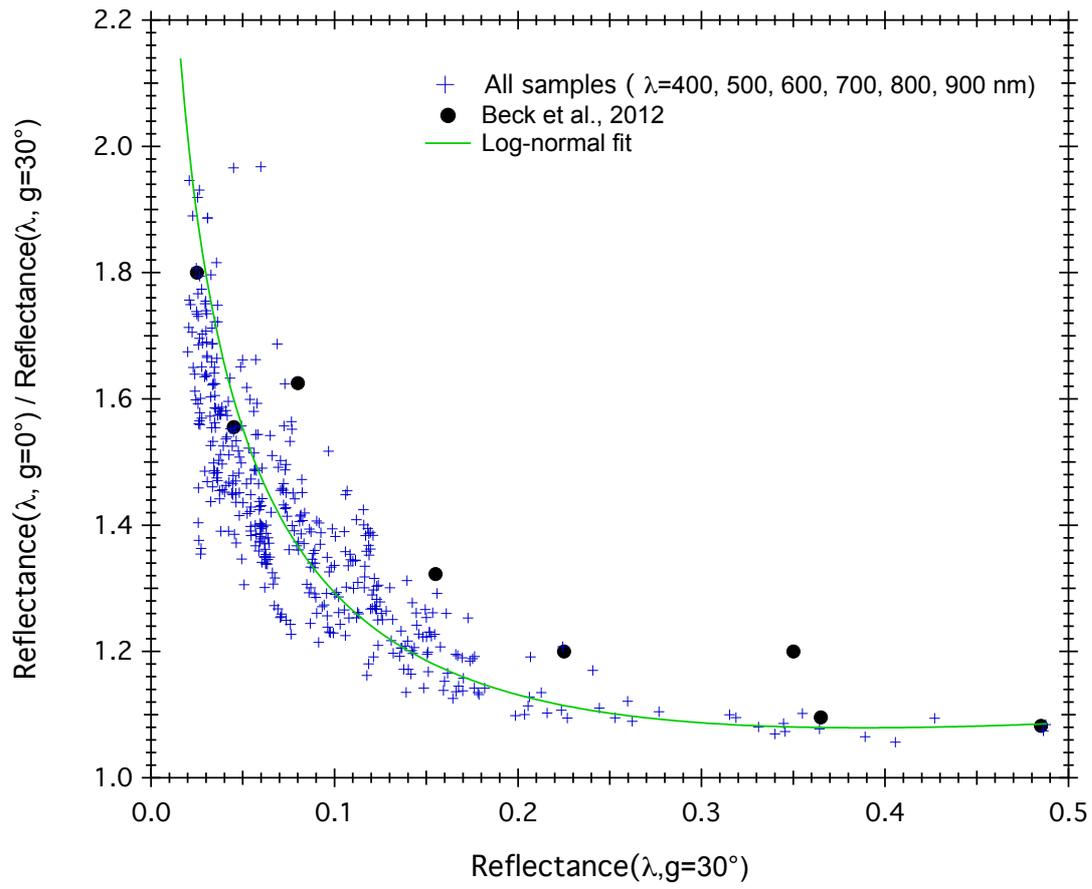

FIGURE 3

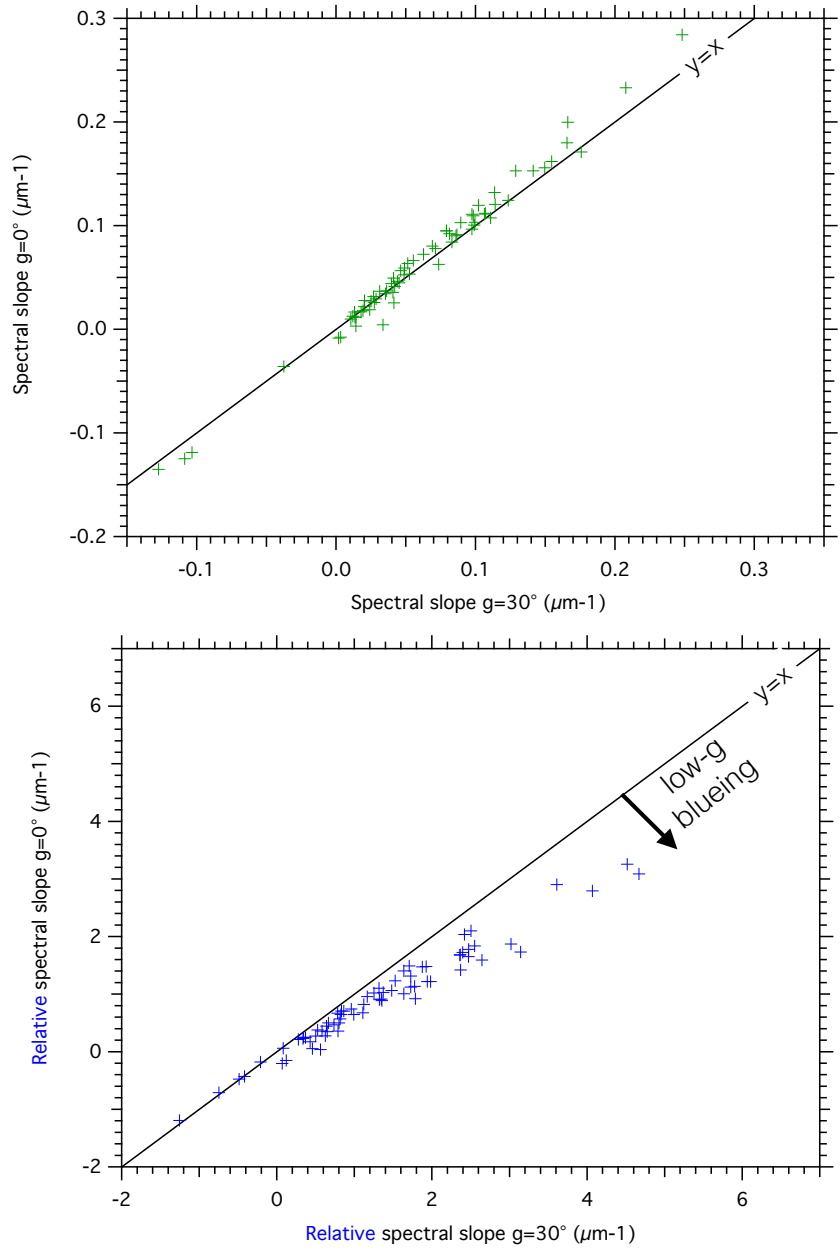

FIGURE 4

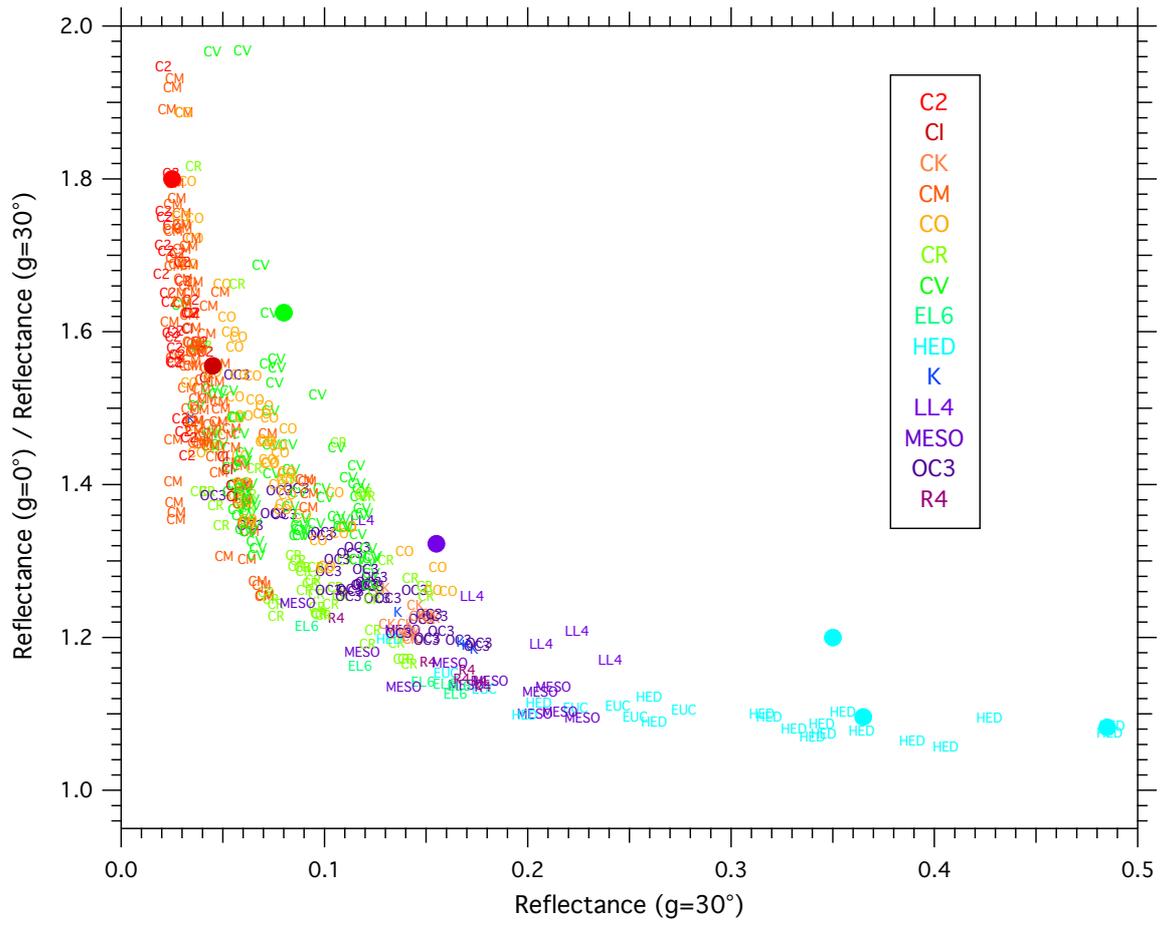

FIGURE 5

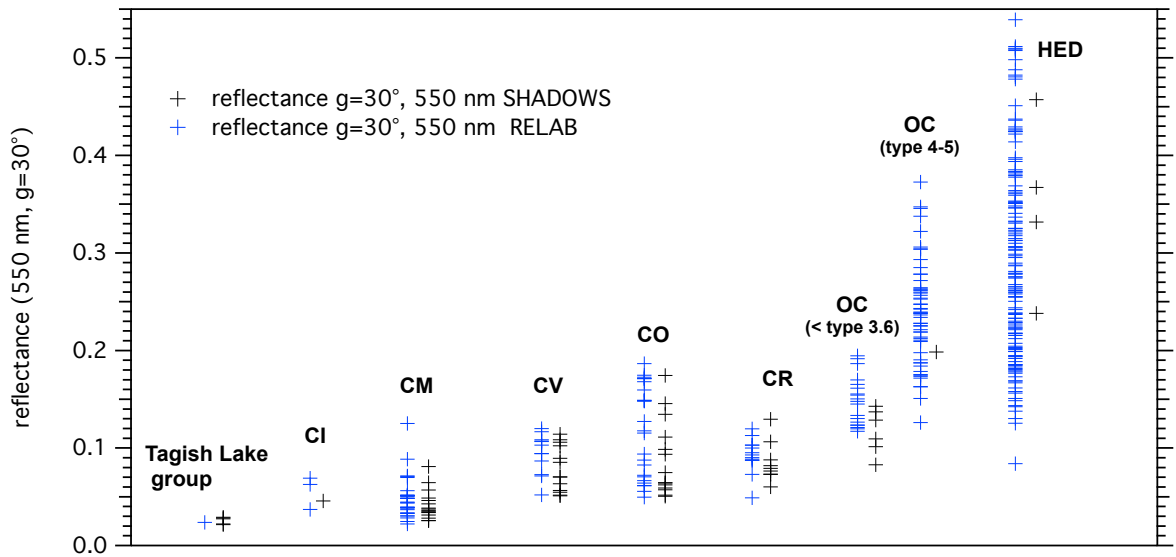

FIGURE 6

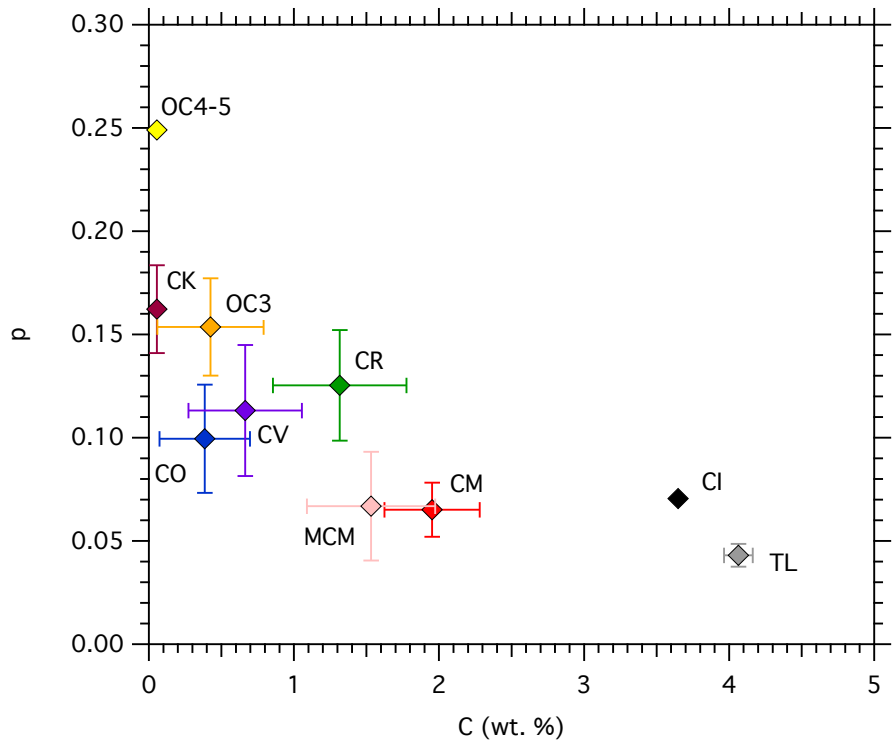

FIGURE 7

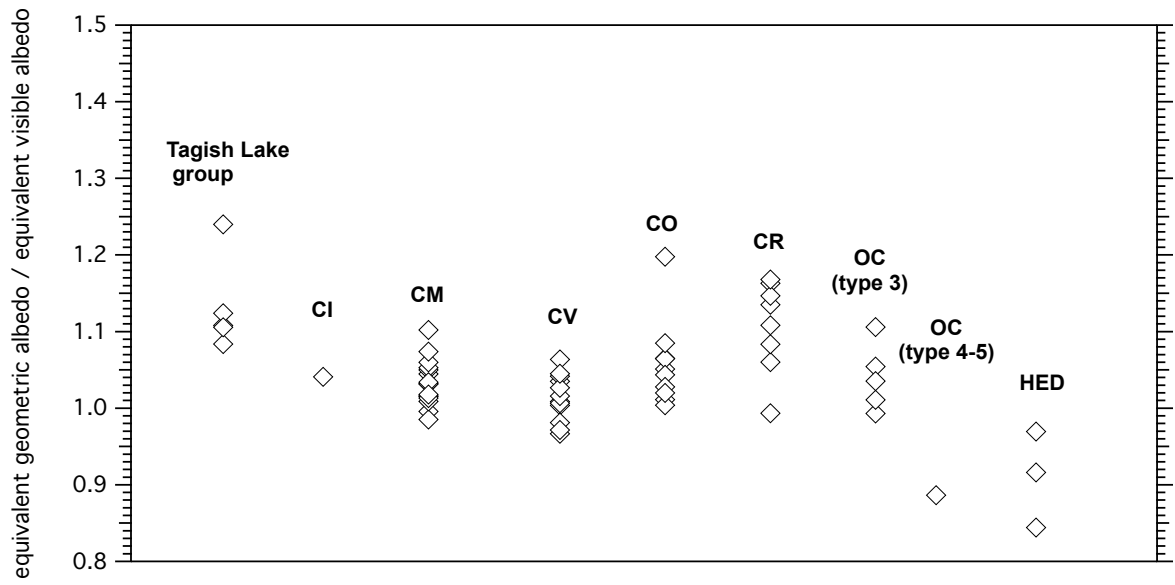

FIGURE 8

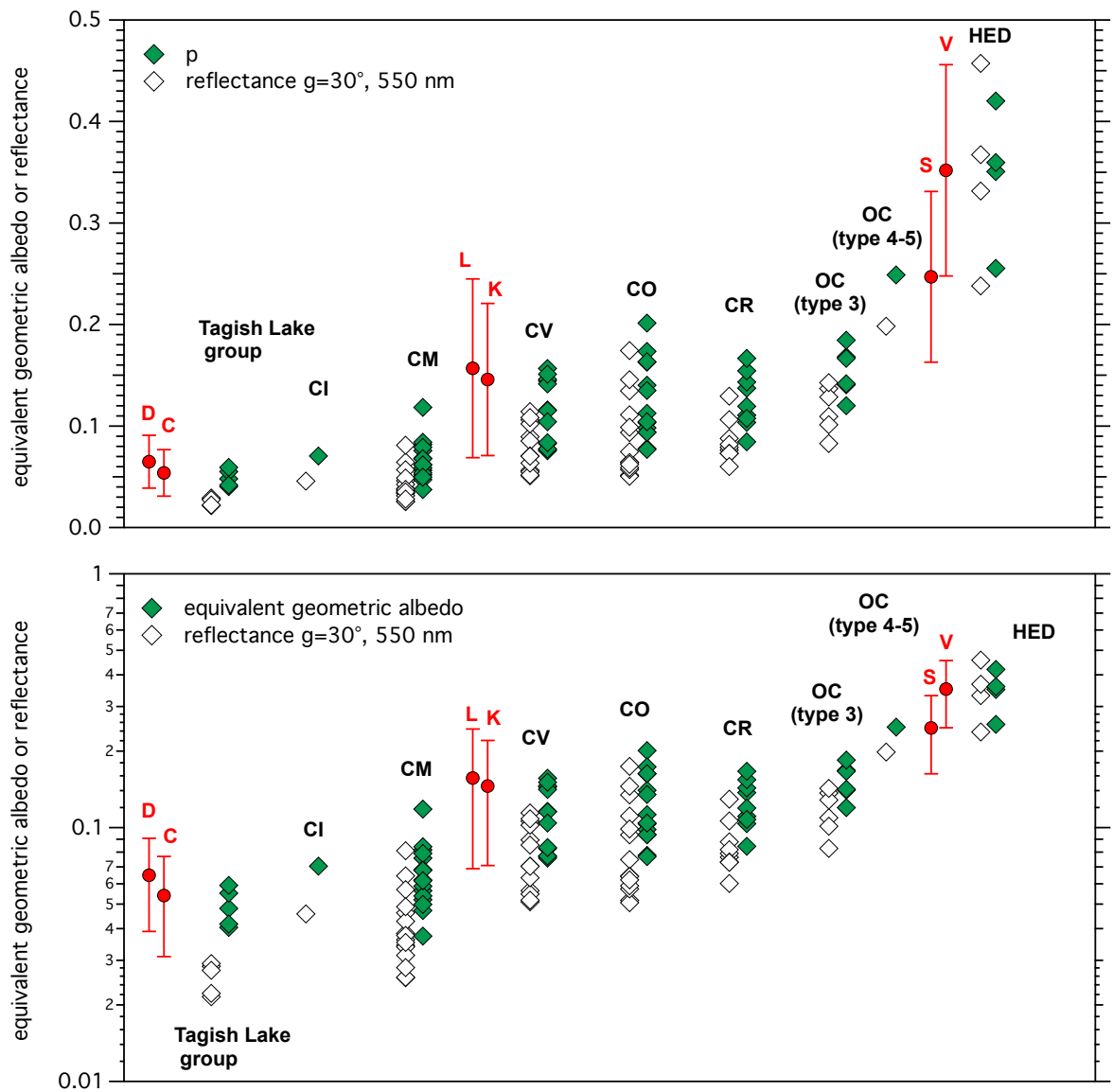

FIGURE 9